\documentclass[11pt,onecolumn]{IEEEtran}
\usepackage{multirow}
\usepackage{setspace}
\usepackage{epsfig}
\usepackage{multicol}
\usepackage{rotating}
\usepackage{tabularx}
\usepackage{url}
\usepackage{enumerate}
\usepackage{graphicx}
\usepackage{amsfonts}
\usepackage{amsmath, amssymb}
\usepackage{caption}
\newcommand{\be}{\begin{eqnarray}}
\newcommand{\ee}{\end{eqnarray}}

\def\tdi{\tilde{d}_i}

\def\eg{{\em e.g.},~}
\def\ie{{\em i.e.},~}

\newcommand{\ben}{\begin{enumerate}}
\newcommand{\een}{\end{enumerate}}
\newcommand{\beq}{\begin{equation}}
\newcommand{\eeq}{\end{equation}}
\newcommand{\beqa}{\begin{eqnarray*}}
\newcommand{\eeqa}{\end{eqnarray*}}
\newcommand{\bit}{\begin{itemize}}
\newcommand{\eit}{\end{itemize}}
\newcommand{\bt}{\begin{tabular}{c}}
\newcommand{\btt}{\begin{tabular}}
\newcommand{\et}{\end{tabular}}

\newcolumntype{P}[1]{>{\centering\arraybackslash}p{#1}}
\newcolumntype{M}[1]{>{\centering\arraybackslash}m{#1}}

\begin{document}

\title{Multicategory Crowdsourcing Accounting for Plurality in Worker Skill and Intention, Task Difficulty, and Task Heterogeneity}

\author{Aditya Kurve\\
\small Department of Electrical Engineering\\
\small The Pennsylvania State University\\
\small \texttt{ack205@psu.edu}\\
\and
David J. Miller\\
\small Department of Electrical Engineering\\
\small The Pennsylvania State University\\
\small \texttt{djmiller@engr.psu.edu}\\
\and
George Kesidis\\
\small Department of Electrical and CSE Engineering\\
\small The Pennsylvania State University\\
\small \texttt{kesidis@engr.psu.edu}\\
}

\maketitle

\begin{abstract}
Crowdsourcing allows to instantly recruit workers on the web to annotate image, web page, or document databases. However, worker unreliability prevents taking a worker's responses at ``face value". Thus, responses from multiple workers are typically aggregated to more reliably infer ground-truth answers. We study two approaches for crowd aggregation on multicategory answer spaces: stochastic modeling-based and deterministic objective function-based. Our stochastic model for answer generation plausibly captures the interplay between worker skills, intentions, and task difficulties and allows us to model a broad range of worker types. Our deterministic objective-based approach does not assume a model for worker response generation. Instead, it aims to maximize the average aggregate confidence of weighted plurality crowd decision making. In both approaches, we explicitly model the \emph{skill} and \emph{intention} of individual workers, which is exploited for improved crowd aggregation. Our methods are applicable in both unsupervised and semisupervised settings, and also when the batch of tasks is \emph{heterogeneous}. As observed experimentally, the proposed methods can defeat ``tyranny of the masses", \ie they are especially advantageous when there is an (\emph{a priori} unknown) minority of skilled workers amongst a large crowd of unskilled (and malicious) workers.
\end{abstract}

\section{Introduction}
Crowdsourcing systems leverage the diverse skill sets of a large number of Internet workers to solve problems and execute projects. In fact, the Linux project and Wikipedia can be considered products of crowdsourcing. These systems have recently gained much popularity with web services such as Amazon MTurk\footnote{www.mturk.com} and Crowd Flower\footnote{www.crowdflower.com}, which provide a systematic, convenient and templatized way for requestors to post problems to a large pool of online workers and get them solved quickly. The success of crowdsourcing has been demonstrated for annotating and labeling images and documents \cite{caltech}, writing and reviewing software code\footnote{www.topcoder.com}, designing products \cite{vukovic2009crowdsourcing}, and also raising funds\footnote{www.crowdfunding.com}. Here, we focus on crowdsourcing tasks with a categorical answer space.

Although the crowd expedites annotation, its anonymity allows noisy or even malicious labeling to occur. Online reputation systems can help reduce the effect of noisy labels, but are susceptible to Sybil \cite{douceur2002sybil} or whitewashing \cite{feldman2004free} attacks. Moreover, the aggregate reputation score only reflects a worker's skill on \emph{previous} tasks/domains. This may not be a good indication of his skill on new domains, for which he has not been evaluated. A second way to mitigate worker unreliability is to assign each task to multiple workers and aggregate their answers in some way to estimate the ground truth answer. The estimation may use simple voting or more sophisticated aggregation methods, \eg \cite{EM-1}, \cite{Karger11_NIPS}, \cite{pasadena}. The aggregation approaches we propose in this work have several notable characteristics. First, we make a clear distinction between worker \emph{skill} level and worker \emph{intention}, \ie whether the worker is honest or malicious. This dissection allows us to plausibly characterize the behavior of an adversarial worker in a multicategory setting. Our aggregation approaches explicitly identify such workers and exploit their behavior to, in fact, improve the crowd's accuracy (relative to the case of non malicious workers). Second, some approaches are only suitable for binary (two choices) tasks and assume that the batch of tasks to be solved all come from the same (classification) domain \cite{pasadena}. By contrast, our approaches explicitly address multicategory tasks and, further do \emph{not} require that all tasks be drawn from the same domain, \ie the batch of tasks can be a heterogeneous mix of problems. We propose two distinct frameworks for jointly estimating workers' skills, intentions, and the ground truth answers for the tasks.

Our first approach assumes a stochastic model for answer generation that plausibly captures the interplay between worker skills, intentions, and task difficulties. To the best of our knowledge, this is the first model that incorporates the difference between worker skill and task difficulty (measured on the real line) in modeling the accuracy of workers on individual tasks. In this model, we formalize the notion of an adversarial worker and discuss and model different types of adversaries. A simple adversary gives incorrect answers ``to the best of his skill level". More ``crafty" adversaries can attempt to evade detection by only giving incorrect answers on the more difficult tasks solvable at their skill level. The detection of adversaries and the estimation of both worker skills and task difficulties can be assisted by the knowledge of ground-truth answers for some (probe) tasks. Unlike some papers such as \cite{taskar1}, we assume that all our probe tasks are labeled by (non-malicious) experts. Accordingly, we formulate a semisupervised approach, invoking a generalized EM (GEM) algorithm \cite{meng1997algorithm} to maximize the joint log likelihood over the (known) true labels for the ``probe" tasks and the answers of the crowd for all tasks. This general approach specializes to an \emph{unsupervised} method when no (labeled) probe tasks are available. Interestingly, our crowd aggregation rule comes precisely from the E-step of our GEM algorithm, since the ground-truth answers are treated as the hidden data \cite{Original-EM} in our GEM approach.

A limitation of the stochastic modeling approach is that, on some domains, its statistical assumptions may not well-characterize crowd behavior. Accordingly, in this work we also investigate ``deterministic" approaches that do not assume underlying stochastic answer generation mechanisms and hence are expected to perform robustly across a diverse set of domains. We propose two deterministic objective-based methods that jointly estimate worker intention, skill and the ground truth answers to multicategory tasks by maximizing a measure of aggregate confidence on the estimated ground truth answers measured over the batch of tasks. Here, crowd aggregation is achieved by weighted plurality voting, with higher weights given to the workers estimated to be the most highly skilled and honest. These workers are identified essentially by their tendency to agree with each other on most of the tasks, unlike the low skilled workers, who tend to answer arbitrarily.

Probe tasks are a form of overhead which limits the number of true tasks in a batch (of fixed size) that the crowd is solving. It may be expensive, time-consuming and/or impractical to \emph{devise} meaningful probe tasks for a given problem domain. Accordingly, we consider our methods in both semisupervised and \emph{unsupervised} settings to evaluate the gains obtained by using probes. Our experimental evaluation of the proposed schemes consisted of three levels. First, we evaluated the schemes using simulated data generated in a way consistent with our proposed stochastic generative model. This allowed us to study the robustness of our GEM algorithm by comparing estimated and actual model parameters. Second, we evaluated and compared performance of the methods using a crowd of ``simulated" workers that do not obviously generate answers in a fashion closely matched to our model. Specifically, each worker was a strong learner, formed as an ensemble of weak learners. Each weak learner was a decision tree, with the ensemble (and thus, a strong learner) obtained by multiclass boosting. A strong worker's skill was controlled by varying the number of boosting stages used. We performed experiments on UC Irvine data sets \cite{UCI} and studied the comparative gains of our methods over benchmark methods on a variety of classification domains. Our final experiment involved a crowdsourcing task we posted using Amazon Mturk. Overall, we observed that our methods are especially advantageous when there is an (\emph{a priori} unknown) minority of skilled workers amongst a large crowd of unskilled (as well as malicious) workers, \ie they are able to overcome ``tyranny of the masses" by estimating the workers' intentions and skills with a reasonable level of accuracy.

\section{Generative Semisupervised Modeling Approach} \label{SS_GEM}
In the following framework, we separately model worker intention and skill. A worker's intention is a binary parameter indicating if he is adversarial or not. An honest worker provides accurate answers ``to the best of his skill level" whereas an adversarial worker may provide incorrect answers ``to the best of his skill level". In the case of binary crowdsourcing tasks, adversarial workers can be identified by a negative weight \cite{Karger11_NIPS} given to their answers. Here we extend malicious/adversarial worker models to \emph{multicategory} tasks and hypothesize both ``simple" and ``crafty" adversaries.

Our approach incorporates task difficulty and worker skill explicitly and, unlike previous approaches \cite{EM-2} \cite{caltech} \cite{Karger11_NIPS}, characterizes the interplay between them. Task difficulty and worker skill are both represented on the real line, with our generative model for a worker's answer based on their difference. If the task difficulty exceeds a worker's skill level, the worker answers randomly (whether honest or adversarial). For an adversary, if the task difficulty is less than his skill level, he chooses randomly only from the set of incorrect answers. We also acknowledge another category of worker type known as ``spammers". These are lazy workers who simply answer randomly for all tasks. Our model well-characterizes such workers via large negative skill values.

\subsection{Notation} \label{notsec}
Suppose a crowd of $N$ workers is presented with a set of $T_u$ unlabeled tasks, for which the ground truth answers are unknown. There are also $T_l$ probe tasks, with known ground truth answers\footnote{The unsupervised setting is a special case where $T_l=0$.}. We assume the crowd is unaware which tasks are probes. Accordingly, a malicious worker cannot alter his answering strategy in a customized way for the probe tasks to ``fool" the system. Let $\{1, 2, ..., T_l\}$ be the index set of the probe tasks and $\{T_l+1, T_l+2, ..., T_l+T_u\}$ be the index set for non-probe tasks. We assume without loss of generality that each worker is asked to solve all the tasks\footnote{We only make this assumption for notational simplicity. Our methodology in fact applies generally to the setting where each worker solves only a subset of the tasks.}. The answers are chosen from a set $\mathcal{C}$ := $\{1, 2, ..., K\}$. Let $z_{i} \in \mathcal{C}$ be the ground truth answer and let $\tdi \in  (-\infty,\infty)$ represent the difficulty level of task $i$. The (ground truth)intention of worker $j$ is indicated by $v_j \in \{0,1\}$, where $v_j = 1$ denotes an honest worker and $v_j=0$ an adversary. $d_j \in (-\infty, \infty)$ represents the $j^{\rm{th}}$ worker's (ground truth) skill level and $a_j$ denotes an additional degree of freedom to introduce variation in the probability mass function across workers (discussed in Section \ref{work}). Finally the response provided to the $i^{\rm{th}}$ task by the $j^{\rm{th}}$ worker is denoted $r_{ij} \in \mathcal{C}$.

Suppose a crowd of $N$ workers is presented with a set of $T_u$ unlabeled tasks, for which the ground truth answers are unknown. There are also $T_l$ probe tasks, with known ground truth answers\footnote{The unsupervised setting is a special case where $T_l=0$.}. We assume the crowd is unaware which tasks are probes. Accordingly, a malicious worker cannot alter his answering strategy in a customized way for the probe tasks to ``fool" the system. Let $\{1, 2, ..., T_l\}$ be the index set of the probe tasks and $\{T_l+1, T_l+2, ..., T_l+T_u\}$ be the index set for non-probe tasks. We assume without loss of generality that each worker is asked to solve all the tasks\footnote{We only make this assumption for notational simplicity. Our methodology in fact applies generally to the setting where each worker solves only a subset of the tasks.}. The answers are chosen from a set $\mathcal{C}$ := $\{1, 2, ..., K\}$. Let $z_{i} \in \mathcal{C}$ be the ground truth answer and let $\tdi \in  (-\infty,\infty)$ represent the difficulty level of task $i$. The (ground truth)intention of worker $j$ is indicated by $v_j \in \{0,1\}$, where $v_j = 1$ denotes an honest worker and $v_j=0$ an adversary. $d_j \in (-\infty, \infty)$ represents the $j^{\rm{th}}$ worker's (ground truth) skill level and $a_j$ denotes an additional degree of freedom to introduce variation in the probability mass function across workers (discussed in Section \ref{work}). Finally the response provided to the $i^{\rm{th}}$ task by the $j^{\rm{th}}$ worker is denoted $r_{ij} \in \mathcal{C}$.

\subsection{Stochastic Generation Model} \label{stoc}
We define our model's \emph{parameter set} as $\Lambda=\{\{(v_j, d_j, a_j)~\forall~j\},\{\tdi ~ \forall~i\}\}$.
We hypothesize the generation of the answers for non-probe tasks in two steps.
Independently for each non-probe task $i \in \{T_l+1,...,T_l+T_u\}$:
\begin{enumerate}
\item Randomly choose the ground truth answer ($z_i$) from $\mathcal{C}$ according to a uniform pmf\footnote{One can always randomize the indexing of the answers for every task to ensure that the true answer is uniformly distributed over $\{1,2,...,K\}$. This would remove any source of bias (\eg toward the true answer being the first ($1$) or the last ($K$)).} $\frac{1}{K}$.
\item For each worker $j\in\{1,...,N\}$, generate $r_{ij}\in \mathcal{C}$ for task $i$ based on the \emph{parameter-conditional} probability mass function (pmf) $\beta(r_{ij}|\Lambda_{ij}, z_i)$, where $\Lambda_{ij}:=\{v_j, d_j, a_j, \tdi\}$\footnote{The specific parametric dependence of $\beta$ on $\Lambda_{ij}$ will be introduced shortly.}.
\end{enumerate}
Also, independently for each probe task $i \in \{1,...,T_l\}$ and each worker $j$, generate the answer $r_{ij}\in \mathcal{C}$  based on the parameter-conditional pmf $\beta(r_{ij}|\Lambda_{ij}, z_i)$.

\subsection{Worker Types}\label{work}
We model the ability of a worker to solve the task correctly using a sigmoid function based on the difference between the task difficulty and the worker's skill\footnote{Alternative (soft) generalized step functions could also in principle be used here.}, \ie the probability that worker $j$ can solve task $i$ correctly is $\frac{1}{1+e^{-a_j(d_j-\tdi)}}$. Note we have included a degree of freedom $a_j$ which attempts to capture the individuality of workers. It is also possible to tie this parameter, \ie set $a_j = a$, $\forall j$.

\subsubsection{Honest Workers}
For an honest worker ($v_j=1$), the pmf $\beta$ is defined as:
\be
\beta(r_{ij}=l|\Lambda_{ij}, v_j=1, z_i) &=& \left\{\begin{array}{ll}
\frac{1}{1+e^{-a_j(d_j-\tdi)}}+\Bigl(\frac{1}{K}\Bigr)\Bigl(\frac{e^{-a_j(d_j-\tdi)}}{1+e^{-a_j(d_j-\tdi)}}\Bigr) & \mbox{for $l=z_i$} \\ \\
\Bigl(\frac{1}{K}\Bigr)\Bigl(\frac{e^{-a_j(d_j-\tdi)}}{1+e^{-a_j(d_j-\tdi)}}\Bigr) & \mbox{otherwise}
\end{array} \right.
\ee
Here, the worker essentially answers correctly with high probability if $d_j>\tdi$, and with probability $\frac{1}{K}$ otherwise. Note that ``spammer" workers, those with $d_j<<\underset{i}{\min}~~\tdi$, will tend to answer randomly for all tasks, under this model. Next, we discuss two models for adversarial workers.

\subsubsection{Simple Adversarial Workers}
For the simple adversarial model, $\beta$ is given by
\be \label{adver}
\beta(r_{ij}=l|\Lambda_{ij}, v_j=0, z_i) &=& \left\{\begin{array}{ll}
\Bigl(\frac{1}{K}\Bigr)\Bigl(\frac{e^{-a_j(d_j-\tdi)}}{1+e^{-a_j(d_j-\tdi)}}\Bigr) &  \mbox{~for $l=z_i$} \\ \\
\Bigl(\frac{1}{K}\Bigr)\Bigl(\frac{e^{-a_j(d_j-\tdi)}}{1+e^{-a_j(d_j-\tdi)}}\Bigr) + \Bigl(\frac{1}{K-1}\Bigr)\Bigl(\frac{1}{1+e^{-a_j(d_j-\tdi)}}\Bigr) & \mbox{~otherwise}
\end{array} \right.
\ee
Here, essentially, the worker only chooses the correct answer (randomly) if the task difficulty defeats his skill level; otherwise he excludes the correct answer and chooses randomly from amongst the remaining answers.

\subsubsection{Complex Adversarial Workers}
In this case, the adversarial worker is more evasive. He answers correctly for simpler tasks with difficulty level below a certain value. Assume $\theta_j < d_j$ to be such a threshold for worker $j$.
The pmf $\beta$ for this (complex) adversarial worker is given by:
\be
\beta(r_{ij}=l|\Lambda_{ij}, v_j=0, z_i) &=& \left\{\begin{array}{ll}
\Bigl(\frac{1}{K}\Bigr)\Bigl(\frac{e^{-a_j(d_j-\tdi)}}{1+e^{-a_j(d_j-\tdi)}}\Bigr) + \Bigl(\frac{1}{K-1}\Bigr)\Bigl(\frac{1}{1+e^{-b_j(\theta_j-\tdi)}}\Bigr)\Bigl(\frac{1}{1+e^{-a_j(d_j-\tdi)}}\Bigr)  &  \\ \mbox{~~~~~~~~~~~~~~~~~~~~~~~~~~~~~~~~~~~~~~~~~~~~~~~~~~~~~~~~~~~~~if $l=z_i$} & \\ \\
\Bigl(\frac{1}{K}\Bigr)\Bigl(\frac{e^{-a_j(d_j-\tdi)}}{1+e^{-a_j(d_j-\tdi)}}\Bigr) + \Bigl(\frac{1}{K-1}\Bigr)\Bigl(\frac{e^{-b_j(\theta_j-\tdi)}}{1+e^{-b_j(\theta_j-\tdi)}}\Bigr)\Bigl(\frac{1}{1+e^{-a_j(d_j-\tdi)}}\Bigr) & \\ \mbox{~~~~~~~~~~~~~~~~~~~~~~~~~~~~~~~~~~~~~~~~~~~~~~~~~~~~~~~~~~~~~~~otherwise} &
\end{array} \right.
\ee

Here, essentially, the worker answers correctly with high probability for easy tasks ($\theta_j > \tdi$), he excludes the correct answer for more difficult tasks below his skill level, and for even more difficult tasks that defeat his skill level ($d_j < \tdi$), he answers correctly at random ($\frac{1}{K}$). \\
In this work we will only experimentally investigate the simple model of adversarial workers.

\subsection{Incomplete, Complete and Expected Complete Data Log Likelihood}
The observed data $\mathcal{X}=\mathcal{R}\cup\mathcal{Z_L}$ consists of the set $\mathcal{R}$ of answers given by the workers to all the tasks, \ie $r_{ij}$ $\forall$ $i,j$ and $\mathcal{Z_L}=\{z_{i}|i \in \{1, 2, ..., T_l\}\}$ is the set of ground truth answers to the probe tasks. We express $\mathcal{R}=\mathcal{R}_L \cup \mathcal{R}_U$, \ie the union of answers to probe tasks and non-probe tasks. We choose the hidden data \cite{Original-EM} to be the ground truth answers to the non-probe tasks, \ie $Z_i$, $i \in \{T_l+1, ..., T_l+T_u\}$. Based on the stochastic model in Section \ref{stoc}, the incomplete data log-likelihood, which we seek to maximize in estimating $\Lambda$, is given by
\be \notag \label{incll}
\mathrm{log} ~\mathcal{L}_{inc}&=& \mathrm{log} ~\mathrm{P}(\mathcal{R},\mathcal{Z}_L|\Lambda)
= \mathrm{log} ~\mathrm{P}(\mathcal{R}_L,\mathcal{R}_U,\mathcal{Z}_L|\Lambda)
= \mathrm{log} ~\mathrm{P}(\mathcal{R}_L,\mathcal{Z}_L|\Lambda)+\mathrm{log} ~\mathrm{P}(\mathcal{R}_U|\Lambda)\\ \notag
&=& \sum_{i=1}^{T_l}\sum_{j=1}^{N}\mathrm{log} ~\frac{1}{K}\beta(r_{ij}|\Lambda_{ij}, z_i)+ \sum_{i=T_l+1}^{T_l+T_u}\sum_{j=1}^{N}\mathrm{log} ~\frac{1}{K}\sum_{k=1}^{K}\beta(r_{ij}|\Lambda_{ij}, Z_i=k)\\
&\propto& \sum_{i=1}^{T_l}\sum_{j=1}^{N}\mathrm{log}~\beta(r_{ij}|\Lambda_{ij}, z_i)+ \sum_{i=T_l+1}^{T_l+T_u}\sum_{j=1}^{N}\mathrm{log} ~\sum_{k=1}^{K}\beta(r_{ij}|\Lambda_{ij}, Z_i=k).
\ee
Treating $Z_i$, $i=1, ..., T_l$ as the hidden data within the EM framework \cite{Original-EM}, the expected complete data log-likelihood, where the expectation is with respect to the pmf $\mathrm{P}(Z_i=k|\mathcal{X},\Lambda)$, can be written as:
\be \notag \label{Expected_ll}
&&\mathbb{E}[\mathrm{log} ~\mathcal{L}_{c}|\mathcal{X},\Lambda]\propto \sum_{i=1}^{T_l}\sum_{j=1}^{N}\mathrm{log} ~\beta(r_{ij}|\Lambda_{ij}, z_i)
+ \sum_{i=T_l+1}^{T_l+T_u}\sum_{j=1}^{N}\sum_{k=1}^{K}\bigl[\mathrm{P}(Z_i=k|\mathcal{X},\Lambda)log~\beta(r_{ij}|\Lambda_{ij}, Z_i=k)\bigr]\\ \notag
&=& \sum_{i=1}^{T_l}\sum_{j:r_{ij}=z_i}\Bigl[v_j~\mathrm{log} \Bigl(\beta(r_{ij}|\Lambda_{ij}, v_j=1, z_i=r_{ij})\Bigr)
+ (1-v_j)~\mathrm{log} \Bigl(\beta(r_{ij}|\Lambda_{ij}, v_j=0, z_i=r_{ij})\Bigr) \Bigr] \\
&+& \sum_{i=1}^{T_l}\sum_{j:r_{ij}\neq z_i} \Bigl[
v_j~\mathrm{log} \Bigl(\beta(r_{ij}|\Lambda_{ij}, v_j=1, z_i\neq r_{ij})\Bigr)
+ (1-v_j)~\mathrm{log} \Bigl(\beta(r_{ij}|\Lambda_{ij}, v_j=0,  z_i \neq r_{ij})\Bigr)\Bigr]
\ee
\vspace{-0.7cm}
\be \notag
+ \sum_{i=T_l+1}^{T_l+T_u}\sum_{k=1}^{K}\sum_{j:r_{ij}=k}\mathrm{P}(Z_i=k)\Bigl[v_j~\mathrm{log} \Bigl(\beta(r_{ij}|\Lambda_{ij}, v_j=1, Z_i=k)\Bigr)
+ (1-v_j)~\mathrm{log} \Bigl(\beta(r_{ij}|\Lambda_{ij}, v_j=0, Z_i=k)\Bigr)\Bigr] \\ \notag
+ \sum_{i=T_l+1}^{T_l+T_u}\sum_{k=1}^{K}\sum_{j:r_{ij}\neq k}\mathrm{P}(Z_i=k)\Bigl[ v_j~\mathrm{log} \Bigl(\beta(r_{ij}|\Lambda_{ij}, v_j=1, Z_i \neq k)\Bigr)
+ (1-v_j)~\mathrm{log} \Bigl(\beta(r_{ij}|\Lambda_{ij}, v_j=0, Z_i \neq k)\Bigr)\Bigr]
\ee
\subsection{The Generalized EM (GEM) Algorithm}
We formulate our algorithm using the above defined expected complete data log-likelihood. The EM algorithm ascends monotonically in $\mathrm{log} ~\mathcal{L}_{inc}$ with each iteration of the E and M steps \cite{Original-EM}. In the \emph{expectation} step, one calculates the pmf $\mathrm{P}(Z_i=k|\mathcal{X},\Lambda^{t})$ using the current parameter values $\Lambda^{t}$, and in the \emph{maximization} step, one computes $\Lambda^{t+1}= \mbox{arg} \max_{\Lambda}~\mathbb{E}[\mathrm{log} ~\mathcal{L}_{c}|\mathcal{X},\Lambda^{t}]$.

\textbf{E step}:
In the E-step we compute the expected value of $\mathcal{Z}_u$ given the observed data $\mathcal{X}$ and the current parameter estimates $\Lambda^{t}$. Based on our assumed stochastic model (section $3.2$), with data for each task generated i.i.d, we have that $\mathrm{P}(\mathcal{Z}_u|\mathcal{X},\Lambda^{t})=\prod_{i=T_l+1}^{T_l+T_u}\mathrm{P}(Z_i=z_i|\mathcal{X},\Lambda^{t})$. Moreover, again based on the assumed stochastic model and applying Bayes' rule, we can derive the closed form expression for the pmf in the E-step as:
\be \label{estep}
\mathrm{P}_i(Z_i=k|\mathcal{X},\Lambda^{t})= \frac{\prod_{j=1}^N\beta(r_{ij}|\Lambda_{ij}^t, Z_i=k)}{\sum_{l=1}^K \prod_{j=1}^N\beta(r_{ij}|\Lambda_{ij}^t, Z_i=l)}, ~~~i\in \{T_l+1,...,T_u+T_l\}.
\ee

\textbf{Generalized M step}:
In the M-step of EM, one maximizes the expected complete data log-likelihood with respect to the model parameters:
\be \label{mst}
\Lambda^{t+1}= arg \max_{\Lambda}~\mathbb{E}[\mathrm{log} ~\mathcal{L}_{c}(\Lambda)|\mathcal{X},\Lambda].
\ee
 Since $\Lambda$ consists of mixed (both continuous and discrete) parameters, with a particular parametric dependence and with $2^{N}$ (honest, adversarial) crowd configurations, it is not practically feasible to find a closed form solution to (\ref{mst}) for our model. Instead, we use a generalized M-step approach \cite{meng1997algorithm}\cite{graham2006unsupervised} to iteratively maximize over the two parameter subsets $\{v_j~\forall~ j\}$, and $\{\{(d_j,a_j)~\forall~j \}, \{\tdi~\forall~i\}\}$.

\textbf{M1 Substep}:
Since (\ref{Expected_ll}) is an additive function of terms that each depend on a single variable $v_j$, we can find a closed form solution for $v_j$ $\forall$ $j$ given all other parameters fixed:
\be
\tilde{v}_j=\underset{v_j\in\{0,1\}}{\text{arg max  }} \mathbb{E} (\mathrm{log}  \mathcal{L}_c(\{\emph{v}_{\emph{j}}\})|\mathcal{X}_j, \tilde{\Lambda} \backslash \{\emph{v}_{\emph{j}}\}).
\ee
Here $\mathcal{X}_j$ is the set of answers provided by the $j^{\rm{th}}$ worker and the ground truth answers for the probe tasks that he answered and $\tilde{\Lambda}$ is the result of the previous M2 substep.

\textbf{M2 Substep}:
We maximize $\mathbb{E}[\mathrm{log} ~\mathcal{L}_{c}(\Lambda \backslash \{\emph{v}_{\emph{j}}\})|\mathcal{X},\{\tilde{\emph{v}}_{\emph{j}}\}]$ with respect to $\Lambda \backslash \{v_j\}$ given $\{\tilde{v}_j\}$ fixed from the previous M1 substep. For this, we use a gradient ascent algorithm which ensures monotonic increase in $\mathrm{log} ~\mathcal{L}_{inc}$, but which may only find a local maximum, rather than a global maximum of $\mathbb{E}[\mathrm{log} ~\mathcal{L}_{c}(\Lambda \backslash \{\emph{v}_{\emph{j}}\})|\mathcal{X},\{\tilde{\emph{v}}_{\emph{j}}\}]$. At convergence, the result is stored in $\tilde{\Lambda}\backslash\{v_j\}$.
The M1 and M2 substeps are applied alternately, iteratively, until convergence. $\Lambda^{t+1}$ stores the result of the generalized M-step at convergence.

\textbf{Inference}:
Note that the E-step (\ref{estep}) computes the \emph{a posteriori} probabilities of ground-truth answers. Thus, after our GEM learning has converged, a maximum \emph{a posteriori} decision rule applied to (\ref{estep}) gives our crowd-aggregated estimates of the true answers for the non-probe tasks.

\subsection{Unsupervised GEM}
Note that when probe tasks are not included in the batch of tasks, an unsupervised specialization of the above GEM algorithm is obtained. In particular, we have $T_l=0$, with the first term in (\ref{incll}) and the first two terms in (\ref{Expected_ll}) not present. Our above GEM algorithm is accordingly specialized for this case. In Section \ref{exptsec}, we will evaluate the unsupervised GEM based scheme along with all other methods.

\section{Energy-Constrained Weighted Plurality Aggregation}
Performance of our GEM approach will in general depend on how well the true answer generation mechanism resembles the one assumed in Section \ref{stoc}. We would also therefore like to explore an alternative ``principle" on which to base crowd aggregation, without any explicit assumption about the underlying stochastic answer generation model. The methods we propose in this section use weighted plurality voting, where the weights assigned to workers essentially reflect their individual skill level. A key idea here is to make the weight vector ``energy-constrained", so as to obtain a bounded solution to the resulting optimization problem. The methods we will propose are applicable to both unsupervised and semisupervised settings, but for clarity of presentation, we will focus on the unsupervised setting and then delineate how to extend these approaches to exploit probe tasks, if available.

\subsection{From simple plurality to weighted plurality voting}
We will introduce and define new variables to clearly explain our approach. Let $T=T_l+T_u$ and $\boldsymbol{\hat{z}}_i$ be the $K \times 1$  vector representing the inferred ground truth answers with $\hat{z}_{im}\in\{0,1\}$ and $\sum_m^K \hat{z}_{im}=1$, \ie $\hat{z}_{im}$ is $1$ when the inferred answer to the $i^{\rm{th}}$ task is $m$. Also, $\boldsymbol{\hat{Z}}=(\boldsymbol{\hat{z}}_i, i=1,...,T)$. All other definitions from section \ref{notsec} will be used. A natural extension of majority voting to the multicategory case is plurality voting, where the answer that gets the maximum number of votes is the inferred answer for that task. To help motivate what follows, we note that plurality voting is the solution of a maximization problem defined over a given batch of tasks. In particular it solves\footnote{Ties could be broken by randomly selecting from among the set of plurality answers.}:
\be
\underset{\boldsymbol{\hat{Z}}}{\text{max  }}\sum_{i=1}^T \sum_{m=1}^K \hat{z}_{im} \sum_{j=1}^N\delta(r_{ij}-m)
\ee
subject to
\be \notag
\hat{z}_{im}\in\{0,1\}, ~~~ \sum_m^K \hat{z}_{im}=1, ~~~ i=1,...,T,
\ee
where $\sum_{j=1}^N\delta(r_{ij}-m)$ is the total vote for answer $i$ by all workers.\\
Plurality-based voting is expected to perform well when all the workers are honest and ``equally reliable", with worker accuracy greater than that of a spammer\footnote{The expected accuracy of the spammer is $\frac{1}{K}$, where $K$ is the number of possible answers for each task.}. However, the crowd may consist of heterogeneous workers with varying skill levels and intentions, as we considered before in Section \ref{SS_GEM}. For the most challenging but realistic ``tyranny of the masses" case where a small proportion of highly skilled workers exist among a mass of unskilled workers or spammers, standard plurality-based voting will be highly sub-optimal. Even supposing the highly skilled workers are \emph{always} correct, ``one worker, one vote" means that, if the skilled worker subset is a small minority, it will not in general be able to ``tip the balance" of the plurality towards the correct answer. Alternatively, here we consider weighted plurality voting schemes, where different weights are assigned to the workers based on their ``accuracy level", accounting for both intention and skill. Allocation of higher weights to the most skilled workers may allow defeating ``tyranny of the masses". Moreover, for weighted plurality voting, ties will almost never occur. In order to ensure well-posed optimization problems, we will impose an energy constraint on the weight vector. We will first propose to jointly estimate the worker weights and the ground truth answers consistent with solving the following optimization problem:
\be \label{opt1}
\underset{\boldsymbol{\hat{Z}},\boldsymbol{w}}{\text{max  }}~~\psi_{\rm{wp}}(\boldsymbol{w},\boldsymbol{\hat{Z}}) = \sum_{j=1}^N w_j\sum_{i=1}^T \sum_{m=1}^K \hat{z}_{im}\delta(r_{ij}-m)
\ee
subject to
\be \notag
\sum_{j=1}^N w_j^2=1,~~\hat{z}_{im}\in\{0,1\}, ~~~ \sum_{m=1}^K \hat{z}_{im}=1.
\ee
Here, we maximize the average weighted plurality score (a measure of aggregate confidence in decisionmaking). (\ref{opt1}) is a non-convex optimization problem, whose (locally optimal) solutions we will obtain via an iterative algorithm, alternating between the updates of the weights and the inferred answers. We iterate over the following two (local maximization) steps until convergence.\\
\textbf{Step 1}: For fixed $\boldsymbol{w}$ and for each task $i$, the choice for  $\boldsymbol{\hat{z}}_i$ which maximizes (\ref{opt1}) is :
\be
\hat{z}_{im}
& = & \left\{\begin{array}{ll}
1 & \mbox{if ${\displaystyle\sum_{j:r_{ij}=m}} w_j  > {\displaystyle\sum_{j:r_{ij}=k}} w_j$ $\forall~k \neq m$}\\
0 & \mbox{otherwise}
\end{array}\right.
\ee
\textbf{Step 2}:
Given fixed $\boldsymbol{\hat{Z}}$ we compute the optimum $\boldsymbol{w}$, maximizing (\ref{opt1}).The Lagrangian for this optimization problem is
    \be
      \mathcal{L}=\sum_{j=1}^N w_j\sum_{i=1}^T \sum_{m=1}^K \hat{z}_{im}(\delta(r_{ij}-m) + \lambda\bigl(\sum_{j=1}^N w_j^2-1\bigr).
    \ee
    Differentiating with respect to $w_k$, we get
    \be
    \frac{\partial\mathcal{L}}{\partial w_k}= \sum_{i=1}^T \sum_{m=1}^K \hat{z}_{im}\delta(r_{ik}-m) + 2\lambda w_k =0 \\ \notag
    \Rightarrow w_k = \frac{-1}{2\lambda}\sum_{i=1}^T \sum_{m=1}^K \hat{z}_{im}\delta(r_{ik}-m)~~~\forall k
    \ee
    We can compute $\lambda$ by squaring the above and then summing over all workers $j$, to find
    \be
    \lambda= -\frac{1}{2}\sqrt{{\displaystyle\sum_{j=1}^N}\Biggl({\displaystyle\sum_{i=1}^T} {\displaystyle\sum_{m=1}^K} v_{im}\delta(r_{ij}-m)\Biggr)^2}
    \ee
    Hence the optimal value of $w_k$ is given by
    \be
     w_k^*= \frac{{\displaystyle\sum_{i=1}^T} {\displaystyle\sum_{m=1}^K} \hat{z}_{im}(\delta(r_{ik}-m)}{\sqrt{{\displaystyle\sum_{j=1}^N}\Biggl({\displaystyle\sum_{i=1}^T} {\displaystyle\sum_{m=1}^K} \hat{z}_{im}\delta(r_{ij}-m)\Biggr)^2}}~~~\forall k.
    \ee

Each of the above two steps ascends in the objective (reward) function $\psi_{\rm{wp}}$, with convergence to a local maximum, starting from an initial weight vector $\boldsymbol{w}=\epsilon\underline{\boldsymbol{1}}$, where $\epsilon=\frac{1}{\sqrt{N}}$ and $\underline{\boldsymbol{1}}$ is an $N \times 1$ vector of all ones.

\subsection{Accounting for Adversaries}
Note that  we have not accounted for adversarial workers in (\ref{opt1}). To do so now, suppose that an adversarial worker $k$ will choose the incorrect answer randomly (uniformly over all incorrect answers). In the following we will develop two extensions of the weight-constrained problem to accommodate the worker's intention. Our first approach uses binary parameters to represent worker intentions, whereas our second method exploits the extra degree of freedom provided by the \emph{sign} of the estimated weight $w_j$ to represent the worker's intention.

\subsubsection{Introducing Binary Parameters to Represent Intention} \label{US_SW}
Suppose we introduce an additional set of variables given by the $N\times1$ vector $\boldsymbol{v}$, where $v_j \in \{0,1\}$ and where $v_j=1$ and $v_j=0$ characterize worker $j$ as honest or adversarial, respectively. Accordingly, we rewrite the optimization problem as:
\be \label{opt2}
\underset{\boldsymbol{\hat{Z}},\boldsymbol{w},\boldsymbol{v}}{\text{max  }}~~\psi_{\rm{bp}}(\boldsymbol{w},\boldsymbol{\hat{Z}},\boldsymbol{v}) = \sum_{j=1}^N\sum_{i=1}^T \sum_{m=1}^K \bigl(v_j\hat{z}_{im}w_j\delta(r_{ij}-m)+ \frac{1}{K-1}(1-v_j)\hat{z}_{im}w_j(1-\delta(r_{ij}-m))\bigr)
\ee
subject to
\be \notag
\sum_{j=1}^N w_j^2=1,~~\hat{z}_{im}\in\{0,1\}, ~~~ \sum_{m=1}^K \hat{z}_{im}=1,~~~v_j\in\{0,1\}.
\ee
Here, when a worker is identified as adversarial, we allocate equal weight ($\frac{w_j}{K-1}$) to all the answers except the one the worker has chosen\footnote{The weights across incorrect answers are normalized by $\frac{1}{K-1}$ so that each worker's contribution to a given task equals his weight $w_j$.}. A locally optimal algorithm, maximizing the objective $\psi_{\rm{bp}}$ starting from an initial weight vector $\boldsymbol{w}=\epsilon\underline{\boldsymbol{1}}$, consists of the following three iterated steps:\\
\textbf{Step 1}: For fixed values of $\boldsymbol{w}$ and $\boldsymbol{v}$ and for each task $i$, the optimal $\boldsymbol{\hat{z}}_i$, maximizing (\ref{opt2}), is chosen as:
\be
\hat{z}_{im}
& = & \left\{\begin{array}{ll}
1 & \mbox{if ${\displaystyle\sum_{j:r_{ij}=m}} v_j w_j + \frac{1}{K-1}{\displaystyle\sum_{j:r_{ij}\neq m}} (1-v_j)w_j > {\displaystyle\sum_{j:r_{ij}=k}} v_jw_j + \frac{1}{K-1}{\displaystyle\sum_{j:r_{ij}\neq k}} (1-v_j)w_j$ $\forall~k \neq m$}\\
0 & \mbox{otherwise}
\end{array}\right.
\ee
\textbf{Step 2}:
For fixed values of $\boldsymbol{\hat{Z}}$ and $\boldsymbol{v}$, the optimal $\boldsymbol{w}$, maximizing (\ref{opt2}), is given by:
    \be \label{optwt1}
     w_k^*= \frac{{\displaystyle\sum_{i=1}^T} {\displaystyle\sum_{m=1}^K} \hat{z}_{im}v_k\delta(r_{ik}-m)+\frac{(1-v_k)}{K-1}(1-\delta(r_{ik}-m))}{\sqrt{{\displaystyle\sum_{j=1}^N}\Biggl({\displaystyle\sum_{i=1}^T} {\displaystyle\sum_{m=1}^K} \hat{z}_{im}v_j\delta(r_{ij}-m)+\frac{(1-v_j)}{K-1}(1-\delta(r_{ij}-m))\Biggr)^2}}~~~\forall~ k
    \ee
\textbf{Step 3}:
For fixed values of $\underline{\boldsymbol{\hat{Z}}}$ and $\boldsymbol{w}$,the optimal $\boldsymbol{v}$, maximizing (\ref{opt2}), is:
\be
v_j
& = & \left\{\begin{array}{ll}
1 & \mbox{if ${\displaystyle\sum_{i=1}^{T}} {\displaystyle\sum_{m=1}^{K}} \hat{z}_{im}w_j\delta(r_{ij}-m) \geq   \frac{1}{K-1}\hat{z}_{im}w_j(1-\delta(r_{ij}-m))$ }~\forall~j\\
0 & \mbox{otherwise}
\end{array}\right.
\ee

\subsubsection{Negative Weights Signify Adversaries} \label{US_COV}
In the previous algorithm, we can see that
\begin{enumerate}[i)]
\item An honest worker's weight contributes to the objective function only if he votes with the (weighted) plurality.
    \item Binary parameters included to represent worker intent result in an additional optimization step. As will be seen in the experiments section, this additional step may be a source of (poor) local optima.
\item The weights computed by (\ref{optwt1}) will always be non-negative, and so the full (real line) range of $w_j$ is not being utilized.
\end{enumerate}
Here, we propose an approach which remedies all three of these ``issues" associated with the previous algorithm. First, let us note that negative weights can be used to signify adversarial workers, and treated accordingly. Thus rather than apportion $\frac{w_j}{K-1}$ when $v_j=0$, as in Section \ref{US_SW}, we can equivalently apportion $\frac{|w_j|}{K-1}$ when $w_j < 0$ (and, as we shall see shortly, avoid the need to explicitly introduce binary intention parameters).
 Second, to appreciate a possible alternative to i), suppose an influential \emph{nonadversarial} worker ($w_j$ large and positive) does not agree with the (current) weighted plurality decision for a given task. Rather than \emph{not} contributing to the plurality score and the objective function, this worker could \emph{subtract} his weight from the weighted plurality score (in fact, from the scores of all answers with which he disagrees) to try to \emph{alter} the weighted plurality decision. Interestingly, we can achieve \emph{both} of these mechanisms, and also avoid including binary intention parameters, with the introduction of a \emph{single} additional cost term, thus modifying the objective function from Section \ref{US_SW}. Specifically, supposing that $\boldsymbol{\hat{Z}}$ is the current estimate of ground truth answers, then worker $j$'s contribution to the objective function is now taken to be:
\be \label{contri}
w_j\sum_{i=1}^T \sum_{m=1}^K \bigl(\hat{z}_{im}\delta(r_{ij}-m)- \frac{1}{K-1}\hat{z}_{im}(1-\delta(r_{ij}-m))\bigr)
\ee
Let us consider the two cases $w_j<0$ and $w_j>0$. If $w_j<0$, then in (\ref{contri}) the adversary's weight magnitude is (fully) subtracted from the current plurality score if his answer agrees with that of the current plurality. On the other hand, if his answer disagrees with the current plurality, his weight magnitude is equally apportioned amongst the remaining answers. Thus, (\ref{contri}) accounts for adversarial workers precisely as we intend. Next, suppose that $w_j>0$. We can see that (\ref{contri}) behaves as desired \emph{in this case as well}, fully adding $w_j$ to the plurality score if he agrees with the plurality decision, and, if he disagrees with the plurality, subtracting $\frac{w_j}{(K-1)}$ from the scores of the answers with which he disagrees. Thus (\ref{contri}) accounts for \emph{both} adversarial and non-adversarial workers precisely in the way we intend. Accordingly, we modify the objective function using (\ref{contri}) as the per-worker contribution:
\be \label{COV}
\underset{\boldsymbol{\hat{Z}},\boldsymbol{w}}{\text{max  }}~~\psi_{\rm{neg}}(\boldsymbol{w},\boldsymbol{\hat{Z}}) = \sum_{j=1}^N w_j\Bigl[\sum_{i=1}^T \sum_{m=1}^K \bigl(\hat{z}_{im}\delta(r_{ij}-m)- \frac{1}{K-1}\hat{z}_{im}(1-\delta(r_{ij}-m))\bigr)\Bigr]
\ee
subject to
\be \notag
\sum_{j=1}^N w_j^2=1,~~\hat{z}_{im}\in\{0,1\}, ~~~ \sum_{m=1}^K \hat{z}_{im}=1.
\ee
We further note that, supposing $\boldsymbol{\hat{Z}}$ are the ground truth answers, then the per worker term bracketed in (\ref{COV}) $\sum_{i=1}^T \sum_{m=1}^K \bigl(\hat{z}_{im}\delta(r_{ij}-m)- \frac{1}{K-1}\hat{z}_{im}(1-\delta(r_{ij}-m))\bigr)$ for a spammer $j$ goes to $0$ as $T\rightarrow\infty$. This follows from the weak law of large numbers and from our assumption that a spammer will randomly choose an answer with a uniform distribution on all possible choices. Consequently, assigning a non-zero weight to a spammer is clearly sub-optimal due to the energy constraint on the weight vector, \ie to maximize (\ref{COV}), spammers will (asymptotic in $T$) be assigned zero weights. Thus (\ref{COV}) properly accounts for honest workers, malicious workers and spammers.

Our locally optimal algorithm for (\ref{COV}) consists of iteration of the following two steps, starting from the weight vector initialization:\\
\textbf{Step 1}: For a fixed $\boldsymbol{w}$ and for each task $i$, choose $\boldsymbol{\hat{z}}_i$ as
\be
\hat{z}_{im}
& = & \left\{\begin{array}{ll}
1 & \mbox{if ${\displaystyle\sum_{j:r_{ij}=m}} w_j - \frac{1}{K-1}{\displaystyle\sum_{j:r_{ij}\neq m}} w_j > {\displaystyle\sum_{j:r_{ij}=k}} w_j - \frac{1}{K-1}{\displaystyle\sum_{j:r_{ij}\neq k}} w_j$ $\forall~k \neq m$}\\
0 & \mbox{otherwise}
\end{array}\right.
\ee
\textbf{Step 2}:
For fixed $\boldsymbol{\hat{Z}}$, the optimal $\boldsymbol{w}$ is:
\be
   w_k^*= \frac{{\displaystyle\sum_{i=1}^T} {\displaystyle\sum_{m=1}^K} \hat{z}_{im}\delta(r_{ik}-m)-\frac{1}{K-1}(1-\delta(r_{ik}-m))}{\sqrt{{\displaystyle\sum_{j=1}^N}\Biggl({\displaystyle\sum_{i=1}^T} {\displaystyle\sum_{m=1}^K} \hat{z}_{im}\delta(r_{ij}-m)-\frac{1}{K-1}(1-\delta(r_{ij}-m))\Biggr)^2}} ~~\forall k.
\ee

\subsection{Semisupervised Case}
Note that for both the methods in Sections \ref{US_SW} and \ref{US_COV}, it is quite straightforward to incorporate probe task supervision. This is achieved by slightly modifying step $1$ in both cases. Specifically, when a task is probe, we simply fix the value of $\boldsymbol{\hat{z}}_i$ to the ground-truth value, rather than using the weighted plurality decision rule.

\section{Experiments} \label{exptsec}
Experiments were performed using synthetic data as well as data generated by a crowdsourced multicategory labeling task on Amazon MTurk. Additionally, for a number of UC Irvine domains, we generated a collection of heterogeneous classifiers to be used as a ``simulated" crowd. We generated adversaries of the simple type in all our experiments. In addition to the proposed schemes, we also compared with simple (multicategory) plurality voting and its semisupervised version, which exploits the probe tasks\footnote{\label{note1}This is weighted plurality, where each worker is weighted by the fraction of probe tasks that he answered correctly.}. Let us reiterate the different methods that we will apply and study in this section:
\begin{enumerate}[i)]
\item PLU: Simple plurality.
\item SS-PLU: Semisupervised plurality.
\item US-SW: Unsupervised objective-based method that uses binary parameter for intention described in Section \ref{US_SW}.
\item SS-SW: Semisupervised specialization of US-SW (using probe tasks).
\item US-NEG: Unsupervised objective-based method without intention parameters, described in Section \ref{US_COV}.
\item SS-NEG: Semisupervised specialization of US-NEG (using probe tasks).
\item SS-GEM: Semisupervised GEM based scheme described in Section \ref{SS_GEM}.
\item US-GEM: Unsupervised specialization of SS-GEM (without any probe tasks).
\end{enumerate}

\subsection{Experiments with Synthetic Data}
\begin{figure}
\centering
  \includegraphics[width=3.5in]{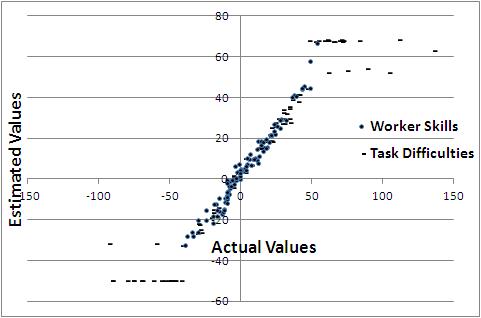}
  \captionof{figure}{Comparison of actual and estimated parameters.}
  \label{fig1}
\end{figure}

\begin{figure}
  \centering
  \includegraphics[width=3.5in]{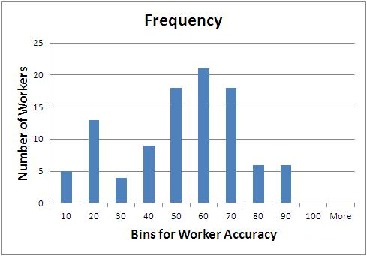}
  \captionof{figure}{Histogram of worker accuracies for high variance regime.}
  \label{fig5}
\end{figure}

These experiments were performed in two parts. For the first part, the synthetic data was produced according to the stochastic generation described in Sections \ref{stoc} and \ref{work}. The goal here was to evaluate the GEM algorithm by comparing the estimated parameters and the estimated hidden ground truth answers with their actual values used in generation of the synthetic data. We generated a crowd of $100$ workers with $d_j \sim \mathcal{N}(1,400)$, $a_j \sim \mathcal{N}(0.3,0.2)$ ; $10\%$ of workers were adversarial. The tasks were generated with $\tdi \sim \mathcal{N}(8,\sigma^2)$, where $\sigma^2$ was varied. The ground truth answer for each task was chosen randomly from $\{0,1,...,4\}$. We observed that in this regime of high variance in worker skill and task difficulty, there is a definite advantage in using the GEM based schemes (SS-GEM and US-GEM) over other schemes, as shown in Table \ref{table1}. Table \ref{table2} shows performance as a function of the number of workers assigned to each task. In each case, a random regular bipartite graph of workers and tasks was generated. We also see in Figure \ref{fig1} the high correlation between the estimated and actual values of worker skills and task difficulties. Also noteworthy from Table \ref{table1} is the superior performance of US-GEM over the other unsupervised objective-based schemes. We plot in Figure \ref{fig5} the histogram of worker accuracies for one of the trials for the first entry in Table \ref{table1}. This illustrates that  highly skilled workers represent a small minority in these experiments.

In the second part we separately evaluated the deterministic objective-based schemes (US-SW and US-NEG) and compared them with simple plurality voting. We performed this for two different models of synthetically generated data. For the first model, all workers belonged to one of three categories: spammers, ``hammers" and adversarial workers.  For all tasks, the spammers answered randomly, the hammers were always correct, whereas the adversarial workers always chose randomly from the incorrect answers. Table \ref{table5} shows comparison across varying proportions of spammers and adversarial workers in the crowd. For the second model, we generated the data according to the stochastic model in Section \ref{stoc} with the task difficulty for each task $\tdi \sim \mathcal{U}[0,8]$. We created three category of workers: high-skilled honest workers ($d_j\sim\mathcal{U}[0,8]$), low-skilled honest workers ($d_j\sim\mathcal{U}[0,2]$), and high-skilled simple adversarial workers ($d_j\sim\mathcal{U}[0,8]$).  The adversarial workers answered incorrectly to the best of their skill level according to (\ref{adver}). Table \ref{table6} shows the comparison of the schemes for this model of synthetic data. Note that for both the experiments in Tables \ref{table5} and \ref{table6}, we averaged the number of erroneous tasks over $20$ trials. We can observe from Tables \ref{table5} and \ref{table6} that US-NEG clearly outperforms US-SW and PLU.

One hypothesis for the performance advantage of US-NEG over US-SW is the additional ``layer" of optimization needed to choose the binary intention parameters in US-SW. This additional layer of optimization could give a greater tendency of US-SW to find (relatively) poor local optimum solutions of its objective function. To test this hypothesis, we devised a hybrid method (US-HYB) which maximizes the US-SW objective function, but starting from $\boldsymbol{\hat{Z}}$ and $\boldsymbol{v}$ initialized based on the US-NEG solution (with the sign of $w_j$ used to initialize $v_j$). We can see from Table \ref{table5} that US-HYB certainly performs better than US-SW. Table \ref{table7} shows the number of trials (out of $100$) when both i) US-HYB gave fewer decision errors than US-SW and ii) the objective function value using US-HYB was strictly greater than using US-SW. We can see from the entries in this table and Table \ref{table5} that there is a strong correlation between US-HYB achieving a greater objective function value and US-HYB achieving fewer decision errors than US-SW.

\begin{table}
\centering
\begin{tabular}{|p{1.7cm}|p{1.5cm}|p{1.5cm}|p{1.5cm}|p{1.5cm}|p{1.5cm}|p{1.5cm}|}
\hline
Task Variance&\multicolumn{6}{c|}{Average erroneous tasks } \\
\cline{2-7}
 &PLU&SS-PLU&US-SW&US-NEG&SS-GEM&US-GEM\\
\hline
4000&23.4&22.1&21.71&22.42&16.9&19.85\\
\hline
2000&21.2&20&20.2&20.5&14.6&15.2\\
\hline
1000&19.2&16.2&18.2&16.5&9.6&9.8\\
\hline
500&11.1&8.1&8.42&7.85&3.14&4.14\\
\hline
250&5.85&5.8&5.28&5.28&2.14&2.57\\
\hline
\end{tabular}
\caption{Synthetic data generated using the stochastic model: Changing task difficulty variance}
\label{table1}
\end{table}

\begin{table}
\centering
\begin{tabular}{|p{1.7cm}|p{1.5cm}|p{1.5cm}|p{1.5cm}|p{1.5cm}|p{1.5cm}|p{1.5cm}|}
\hline
Assignment degree&\multicolumn{6}{c|}{Average erroneous tasks } \\
\cline{2-7}
 &PLU&SS-PLU&US-SW&US-NEG&SS-GEM&US-GEM\\
\hline
20&31.28&31.6&31.2&29.85&27&28\\
\hline
40&22.85&21.2&21.42&21.14&18.42&18.85\\
\hline
60&19.42&18.6&18&18&14.28&15.42\\
\hline
80&16.57&16&15.4&15.14&14.14&13.4\\
\hline
\end{tabular}
\caption{Synthetic data generated using the stochastic model: Changing number of worker assignments.}
\label{table2}
\end{table}

\begin{table}
\centering
\begin{tabular}{|p{2cm}|p{1.5cm}|p{1.5cm}|p{1.5cm}|p{1.5cm}|p{1.5cm}|p{1.5cm}|}
\hline
Percentage adversarial workers&\multicolumn{6}{c|}{\shortstack{Number of erroneous tasks\\US-SW/US-NEG/\\US-HYB/PLU}} \\
\cline{2-7}
 &\multicolumn{6}{c|}{Percentage spammers}\\
\cline{2-7}
 &10&20&30&40&50&60\\
\hline
0&5.4/5.1/\newline 5.2/6.9&6.7/5.9/\newline 6/8.7&6.0/5.7/\newline 5.6/9.6&10.5/9.05/\newline 9.1/16.2&10.9/9.5/\newline 9.7/16.6&13.7/11.7/\newline 11.4/22.3\\
\hline
5&5.6/5.4/\newline 5.3/8.1&6.6/6.3/\newline 6.4/9.8&8.7/8.6/\newline 8/13.1&10.1/9.3/\newline 9.4/16.9&12.8/11.9/\newline 12.0/20.8&19.2/17.1/\newline 16.9/29.5\\
\hline
10&6.8/6.1/\newline 6.0/10.4&7.3/6.2/\newline 6.6/11.9&8.8/8.0/\newline 8.5/15.8&10.3/9.4/\newline 9.5/18.3&13.8/12.4/\newline 12.3/24&29.7/16.2/\newline 16.3/30.7\\
\hline
15&5.8/5.0/\newline 5.0/9.7&8.4/7.5/\newline 7.7/14.4&10.9/9.8/\newline 9.9/18.1&13.5/12.5/\newline
12.7/23.4&16.6/14.8/\newline 15.5/30.2&24.3/21.0/\newline 21.3/37.8\\
\hline
20&8.1/7.1/\newline 7.5/13.8&8.3/3.7/\newline 8.5/16.6&12.2/11.2/\newline 11.3/22.2&13.7/11.8/\newline 12.2/27&18.8/16.8/\newline 16.7/34.2&29.2/22.3/\newline 22.4/44.8\\
\hline
\end{tabular}
\caption{Synthetic data: Comparing objective-based methods for spammer, hammer and adversary model}
\label{table5}
\end{table}

\begin{table}
\centering
\begin{tabular}{|p{2cm}|p{1.5cm}|p{1.5cm}|p{1.5cm}|p{1.5cm}|p{1.5cm}|p{1.5cm}|}
\hline
Percentage high skilled adversarial workers&\multicolumn{6}{c|}{\shortstack{Number of erroneous tasks\\PLU/US-SW/\\US-NEG}} \\
\cline{2-7}
 &\multicolumn{6}{c|}{Percentage low skilled honest workers}\\
\cline{2-7}
 &10&20&30&40&50&60\\
\hline
5&9.4/6.7/\newline 6.0&10.5/6.5/\newline 6.1&12.7/6.9/\newline 6.7&14.4/8.3/\newline 7.2&20.3/11.3/\newline 10.5&26.9/17.3/\newline 15.4\\
\hline
10&9.8/6.9/\newline 6.1&12.1/6.4/\newline 6.1&15.8/8.6/\newline 7.7&18.4/10.5/\newline 9.8&23.1/13.2/\newline 12.6&29.5/18.3/\newline 16\\
\hline
15&10.5/6.0/\newline 5.6&13.5/8.2/\newline 7.1&16.7/9.4/\newline 7.7&22.9/12.8/\newline 11.3&27.7/14.5/\newline 13.7&37/22.4/\newline 20.1\\
\hline
20&12.8/6.7/\newline 6.4&17.4/9.7/\newline 8.5&20.4/9.1/\newline 8.4&25.7/14.7/\newline 12.6&37.4/20.3/\newline 18.2&42.5/27.4/\newline 22.4\\
\hline
\end{tabular}
\caption{Synthetic data: Comparing objective-based methods for data generated using model in Section \ref{stoc}}
\label{table6}
\end{table}

\begin{table}
\centering
\begin{tabular}{|p{2cm}|p{1.8cm}|p{1.8cm}|p{1.8cm}|p{1.8cm}|p{1.8cm}|p{1.8cm}|}
\hline
Percentage Adversarial&\multicolumn{6}{c|}{Number of trials} \\
\cline{2-7}
 &\multicolumn{6}{c|}{Percentage Spammers}\\
\cline{2-7}
 &10&20&30&40&50&60\\
\hline
0&34&42&34&67&63&45\\
\hline
5&17&47&49&54&62&76\\
\hline
10&24&57&48&49&64&77\\
\hline
15&71&31&62&57&81&95\\
\hline
20&55&41&45&66&91&94\\
\hline
\end{tabular}
\caption{Comparing US-SW and US-HYB}
\label{table7}
\end{table}

\subsection{Simulating a Crowd using an Ensemble of Classifiers}
We also leveraged ensemble classification to generate a set of automated workers (each an ensemble classifier) using boosting \cite{schapire1990strength}. Each such classifier (worker) is a strong learner obtained by applying multiclass boosting to boost decision tree-based weak learners. The strength (accuracy) of each worker was varied by controlling the number of boosting stages. Each weak learner was trained using a  random subset of the training data to add more heterogeneity across the workers' hypotheses. Note that unlike Section $5.1$, this approach to simulated generation of a crowd is \emph{not} obviously matched to our stochastic data generation model in Section \ref{stoc}. Thus, this more complex simulation setting provides a more realistic challenge for our model and learning.

We ran Multiboost \cite{benbouzid2012multiboost} $100$ times to create a crowd of $100$ workers for four domains that are realistic as crowdsourcing domains: Pen Digits\footnote{We resampled the dataset to have only odd digits.}, Vowel, Dermatology, and Nominal\footnote{Hungarian named entity dataset \cite{szarvas2006multilingual}. Identifying and classifying proper nouns into four categories: not a proper noun, person, place, and organization.}. For each experimental trial, $100$ crowd tasks were created by randomly choosing $100$ data samples from a given domain; $10$ of them were randomly chosen to be probe. The rest of the data samples from the domain were used for training the strong (ensemble) classifiers/workers. The average of the number of crowd-aggregated erroneous tasks was computed across $5$ trials, where each trial consisted of a freshly generated crowd of workers and set of tasks. In Tables \ref{table3}, \ref{table8}, \ref{table9}, and \ref{table10} we give performance evaluation for different worker accuracy means and variances for the four domains. We did not directly control these values since they were an outcome of the boosting mechanism. However we could control the number of boosting stages used by each worker. We also show the performance when $10\%$ of the workers were replaced by adversarial workers. These synthetic adversaries retained the estimated skill level of the (replaced) workers and generated their answers using the stochastic model described in section $3.2$. In Table \ref{table3}, the worker accuracy mean and variance across all workers is based on the number of correctly answered tasks (both probe and non-probe) for each worker computed in the absence of adversaries.

\begin{figure}
  \centering
  \includegraphics[width=3.5in]{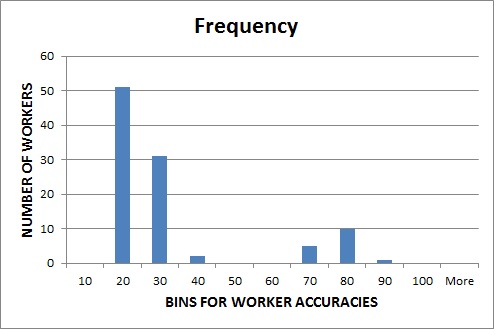}
  \captionof{figure}{Pen Digits dataset: Histogram of worker accuracies with a skewed distribution of skills}
  \label{pendigit1}
\end{figure}

\begin{figure}
  \centering
  \includegraphics[width=3.5in]{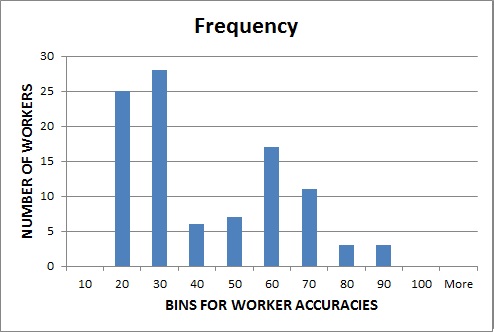}
  \captionof{figure}{Pen Digits dataset: Histogram of worker accuracies with a more gradual ``spread" of skill distribution}
  \label{pendigit2}
\end{figure}

\begin{figure}
  \centering
  \includegraphics[width=3.5in]{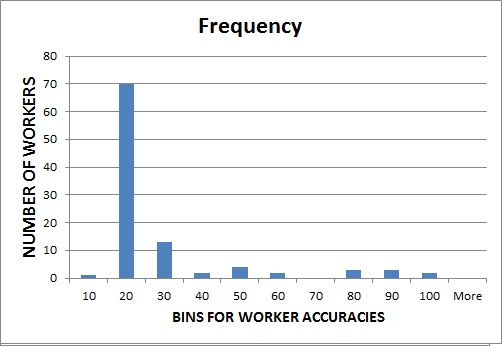}
  \captionof{figure}{Dermatology dataset: Histogram of worker accuracies. Supervision greatly helps}
  \label{Derma1}
\end{figure}

\begin{figure}
  \centering
  \includegraphics[width=3.5in]{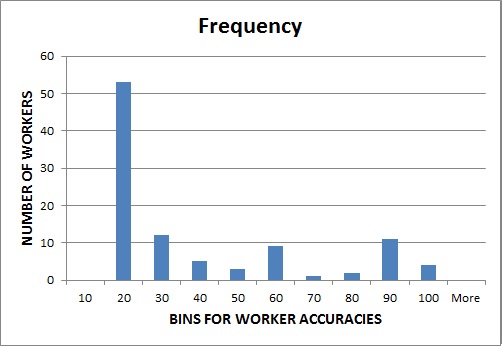}
  \captionof{figure}{Dermatology dataset: Histogram of worker accuracies. Supervision does not help much.}
  \label{Derma2}
\end{figure}

\begin{figure}
  \centering
  \includegraphics[width=3.5in]{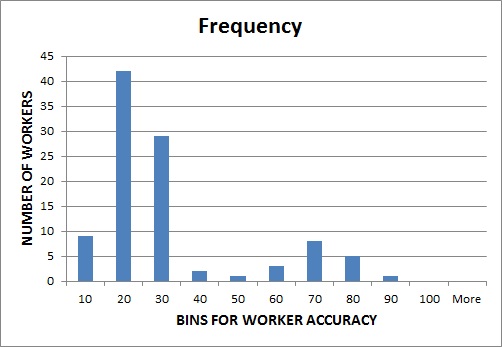}
  \captionof{figure}{Vowel dataset: Histogram of worker accuracies.}
  \label{vowel1}
\end{figure}

We can see in Table \ref{table3} for the Pen Digits dataset, the gain in performance with our methods, especially in the presence of adversarial workers. Note that for low means of worker accuracy, the GEM based methods outperform others, whereas the weighted plurality based methods using negative weights for adversaries (US-NEG and SS-NEG) perform better than other schemes for relatively higher means of worker accuracy. Figures \ref{pendigit1} and \ref{pendigit2} show the histogram of worker accuracies for low and high mean cases, respectively. In Figure \ref{pendigit1}, we plot the histogram of worker accuracies corresponding to the first entry in Table \ref{table3} (for a single trial). We can observe an extremely skewed distribution of worker accuracies where a tiny minority of extremely high skilled workers exist amidst others who are mostly spammers. This distribution seems to best ``agree"  with the GEM based schemes. Figure \ref{pendigit2} corresponds to the last entry in Table \ref{table3}. Here we observe a less skewed distribution of worker accuracies. The objective based US-NEG and SS-NEG schemes perform better in this regime. The GEM based scheme is able to exploit simple adversaries much more than other methods, achieving improved crowd aggregation accuracy compared to the case where no adversaries are present, for almost all the entries in Table \ref{table3}. Also observed from the table on this dataset, $10$\% probe task supervision does not greatly improve performance, \eg US-GEM performs as well as SS-GEM. Moreover, SS-PLU is highly sub-optimal when probe tasks are available. From these results, we can see that our proposed schemes are able to identify and leverage the expertise of a small subset of highly skilled workers, thus defeating ``tyranny of the masses". As seen in Table \ref{table8} on Dermatology, although unsupervised, US-NEG performs very close to SS-GEM when adversarial workers are absent. However, its performance degrades when adversaries are introduced in the crowd. SS-GEM greatly outperforms other methods. Also note that, on this datset, unlike Pen Digits, probe supervision greatly assists the inference using GEM for lower mean worker accuracy (as seen in the first two entries of Table \ref{table8}). In Figures \ref{Derma1} and \ref{Derma2}, we plot the worker accuracy profiles for two cases: where probe supervision is and is not greatly beneficial, respectively. Table \ref{table9} shows the results on the vowel dataset. Here, US-NEG and SS-NEG clearly outperform all other methods, even SS-GEM, in the absence of adversaries. US-GEM and SS-GEM, again, perform better than all others when adversarial workers are introduced. Figure \ref{vowel1} shows the distribution of worker accuracies in the crowd in a histogram for one of the trials corresponding to the lowest mean accuracy in Table \ref{table9}. We observed from our experiments with the Nominal dataset (Table \ref{table10}) that the GEM based methods perform better than others overall, although the performance gap diminishes as the mean worker accuracy improves.

\begin{table}
\footnotesize
\begin{tabular}{|p{1.5cm}|p{1.5cm}|p{1.5cm}|p{1cm}|p{1cm}|p{1cm}|p{1cm}|p{1cm}|p{1cm}|p{1cm}|p{1cm}|}
\hline
Worker accuracy mean&Worker accuracy variance&Task accuracy variance& \multicolumn{8}{M{8cm}|}{\shortstack{\\Without adversarial workers\\With $10$\% adversarial workers}}\\
\cline{4-11}
& & &\tiny{PLU}&\tiny{SS-PLU}&\tiny{US-SW}&\tiny{SS-SW}&\tiny{US-NEG}&\tiny{SS-NEG}&\tiny{US-GEM}&\tiny{SS-GEM}\\
\hline
24.19&262.8&24.23&23.8\newline45.2&11\newline13.8&11.2\newline14.2&11.1\newline14.2&10.8\newline10.8&10.8\newline10.7&9.4\newline5.2&8.4\newline5.2\\
\hline
26.27&294.2&32.5&22\newline29&10.4\newline13.8&11\newline13.8&11\newline13.8&12\newline12&11.8\newline12&8.6\newline5.4&8.6\newline5.3\\
\hline
29.5&465.5&42.9&17.6\newline21.4&11.8\newline14.4&12\newline15&11.9\newline14.9&10.8\newline13&10.8\newline13&9.5\newline6.4&9.2\newline5.8\\
\hline
33.9&485.6&88.8&14.8\newline17.2&8.8\newline10.2&9.4\newline10.8&9.4\newline10.8&7.2\newline7.2&7\newline7.2&9.6\newline8&9\newline7.4\\
\hline
38.8&522.6&182.2&14\newline16.4&7.6\newline7.8&7.8\newline8.2&7.8\newline8.2&6\newline6.6&6\newline6.6&8.2\newline8.2&7.8\newline8\\
\hline
\end{tabular}
\caption{Experiments using Pen Digits dataset}
\label{table3}
\end{table}

\begin{table}
\footnotesize
\begin{tabular}{|p{1.5cm}|p{1.5cm}|p{1.5cm}|p{1cm}|p{1cm}|p{1cm}|p{1cm}|p{1cm}|p{1cm}|p{1cm}|p{1cm}|}
\hline
Worker accuracy mean&Worker accuracy variance&Task accuracy variance& \multicolumn{8}{M{8cm}|}{\shortstack{\\Without adversarial workers\\With $10$\% adversarial workers}}\\
\cline{4-11}
& & &\tiny{PLU}&\tiny{SS-PLU}&\tiny{US-SW}&\tiny{SS-SW}&\tiny{US-NEG}&\tiny{SS-NEG}&\tiny{US-GEM}&\tiny{SS-GEM}\\
\hline
19.7&215.1&44.1&48\newline62.4&23.2\newline23&23\newline45&22.9\newline45&15.8\newline50.8&15.8\newline50.8&24.2\newline32.4&8.6\newline7.4\\
\hline
22.6&399.8&66.3&40.2\newline50.4&12.2\newline11.2&13\newline24.6&12.8\newline24.2&6\newline19.2&6\newline19.2&20.4\newline17.4&8.2\newline7\\
\hline
26.9&480.3&76.6&22.4\newline30&8.6\newline8.4&8.4\newline8.6&8.3\newline8.6&7\newline23.2&6.8\newline23.1&7\newline4.6&6.8\newline4\\
\hline
27.6&563.1&88.9&26.8\newline36.6&9.6\newline9&9.8\newline9.8&9.7\newline9.8&6.8\newline14&6.8\newline14&5.6\newline2.8&5.4\newline2.2\\
\hline
31.7&632.1&110.3&14.4\newline19&4\newline5&4.4\newline4&4.4\newline4&4.8\newline4.3&4.8\newline4.2&4\newline3.8&4\newline3.6\\
\hline
\end{tabular}
\caption{Experiments using Dermatology dataset}
\label{table8}
\end{table}

\begin{table}
\footnotesize
\begin{tabular}{|p{1.5cm}|p{1.5cm}|p{1.5cm}|p{1cm}|p{1cm}|p{1cm}|p{1cm}|p{1cm}|p{1cm}|p{1cm}|p{1cm}|}
\hline
Worker accuracy mean&Worker accuracy variance&Task accuracy variance& \multicolumn{8}{M{8cm}|}{\shortstack{\\Without adversarial workers\\With $10$\% adversarial workers}}\\
\cline{4-11}
& & &\tiny{PLU}&\tiny{SS-PLU}&\tiny{US-SW}&\tiny{SS-SW}&\tiny{US-NEG}&\tiny{SS-NEG}&\tiny{US-GEM}&\tiny{SS-GEM}\\
\hline
24.2&390.1&35.6&25.6\newline26.2&15.8\newline15.6&15.8\newline15.2&15.8\newline15.2&13.4\newline14.2&13.3\newline14.1&16.2\newline11.8&15.6\newline10\\
\hline
26.8&412.4&42.3&23.1\newline26&14.2\newline14.7&14.7\newline15.4&14.7\newline15.4&13.4\newline12.1&13.2\newline12&15.9\newline10.4&15.7\newline9.4\\
\hline
29&445.8&57.4&22.5\newline42.4&13.8\newline14.4&13.9\newline13.6&13.9\newline13.3&12.8\newline11.4&12.8\newline11.4&15.6\newline9.6&15.2\newline5.2\\
\hline
32.2&624.6&81.8&17.8\newline25&12.2\newline13&12.4\newline12&12.4\newline11.8&11\newline13.8&10.8\newline13.5&13.6\newline7.8&13.6\newline7.6\\
\hline
35.9&716.4&118.9&16.4\newline20&13.6\newline13.4&13.4\newline12.8&13.3\newline12.8&13.2\newline11.8&13.2\newline11.7&16.4\newline12.2&15.6\newline9.2\\
\hline
\end{tabular}
\caption{Experiments using Vowel dataset}
\label{table9}
\end{table}

\begin{table}
\footnotesize
\begin{tabular}{|p{1.5cm}|p{1.5cm}|p{1.5cm}|p{1cm}|p{1cm}|p{1cm}|p{1cm}|p{1cm}|p{1cm}|p{1cm}|p{1cm}|}
\hline
Worker accuracy mean&Worker accuracy variance&Task accuracy variance& \multicolumn{8}{M{8cm}|}{\shortstack{\\Without adversarial workers\\With $10$\% adversarial workers}}\\
\cline{4-11}
& & &\tiny{PLU}&\tiny{SS-PLU}&\tiny{US-SW}&\tiny{SS-SW}&\tiny{US-NEG}&\tiny{SS-NEG}&\tiny{US-GEM}&\tiny{SS-GEM}\\
\hline
45.4&575.4&339.1&19\newline22.6&18.8\newline18.8&18.8\newline18.8&18.8\newline18.8&18.2\newline18.3&18.2\newline18.3&18.4\newline18.1&18.2\newline18.1\\
\hline
49.1.3&599.2&328.6&10.4\newline11.6&8.4\newline8.2&9.2\newline9.4&9.1\newline9.4&8.1\newline8.1&8.1\newline7.8&7.7\newline7&6.1\newline5.9\\
\hline
51.3&649.9&325.1&9.8\newline12.2&7.8\newline6.6&8\newline8.4&8\newline8.4&7.8\newline7.8&7.6\newline7.7&7.7\newline6.1&5.6\newline5.4\\
\hline
52.6&678.5&298.5&9.6\newline11&6.2\newline4.8&6.7\newline6.5&6.7\newline6.4&5.3\newline5.2&5.3\newline5.2&5.1\newline3.2&4.7\newline2.9\\
\hline
54.3&878.1&233.1&4.2\newline5.2&3.2\newline3&3.2\newline3.2&3.1\newline3.2&3.2\newline3.2&3.2\newline3.2&2.9\newline2.4&2.6\newline2.3\\
\hline

\end{tabular}
\caption{Experiments using Nominal dataset}
\label{table10}
\end{table}

\subsection{MTurk Experiment}
\begin{figure}[h]
\centering
\includegraphics[width=\linewidth]{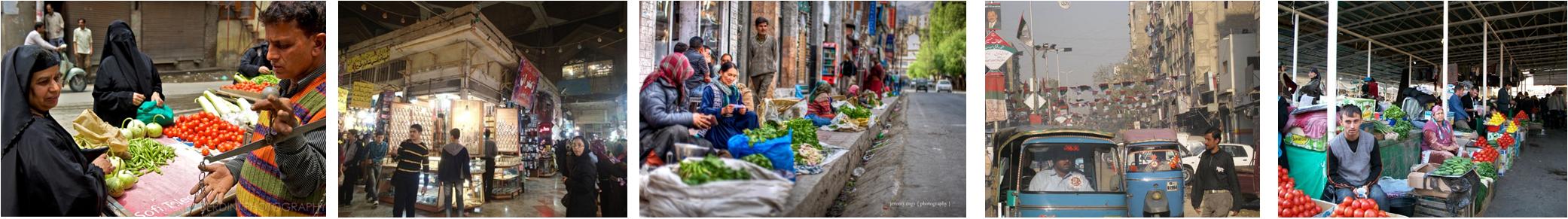}
\caption{The MTurk experiment: a few of the sample images}
\label{fig3}
\end{figure}
\begin{figure}
  \centering
  \includegraphics[width=3.5in]{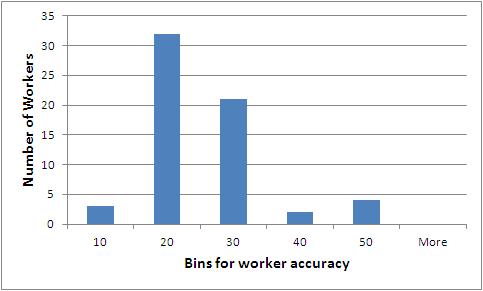}
  \captionof{figure}{Histogram of worker accuracies for the MTurk experiment.}
  \label{fig4}
\end{figure}

We designed an image labeling task where workers had to provide the country of origin, choosing from \emph{Afghanistan, India, Iran, Pakistan, Tajikistan}. Some of the regions in these countries look very similar in their culture, geography, and demography and hence only people with domain experience and a deep understanding of the region will likely know the true answer. For instance, the blue rickshaws are typical to Pakistan and the yellow taxis are more common in Kabul. One can also guess \emph{e.g.} from the car models on the street or from the script on street banners. We posted a task consisting of $50$ such images on Amazon MTurk and asked all workers to upload a file with their answers on all tasks. We received responses from $62$ workers. In order to evaluate under the scenario where workers answer only a subset of the tasks, for each task, we used answers from a sampled set of workers using a randomly generated degree regular bipartite graph consisting of worker and task nodes. A worker's answer to a task was used only when a link existed in the bipartite graph between the two corresponding nodes. Table \ref{table4} shows the average number of erroneous crowd-aggregated answers for the methods under study as we varied the number of tasks assigned to each worker. The average was computed over $5$ trials, each time generating a new instance of a random bipartite graph and using $5$ randomly chosen probe tasks. The histogram of worker accuracies is shown in Figure \ref{fig4}. From the histogram we can observe that only a small fraction of workers could answer most tasks correctly.

\begin{table}
\centering
\footnotesize
\begin{tabular}{|p{2.5cm}|p{1.3cm}|p{1.3cm}|p{1.3cm}|p{1.3cm}|p{1.3cm}|p{1.3cm}|p{1.3cm}|p{1.3cm}|}
\hline
Assignment Degree&PLU&SS-PLU&US-SW&SS-SW&US-NEG&SS-SW&US-GEM&SS-GEM \\
\hline
10&15.4&18.2&15.1&15.1&14.3&14.3&13.9&13.4\\
\hline
20&14.2&13.6&14.1&14.1&13.8&13.7&7.2&7.2\\
\hline
30&14.2&11.2&13.5&13.4&13.2&13.2&5.2&5.2\\
\hline
40&12.2&7.6&11.5&11.5&10.1&10.1&4.2&4\\
\hline
50&13.4&8&13.2&13.2&12.5&12.5&3.4&3.4\\
\hline
\end{tabular}
\caption{MTurk Experiment: Average number of erroneous tasks}
\label{table4}
\end{table}

\section{Related Work}
Crowdsourcing has attracted much interest from the research community and it has been studied in different forms and under various scenarios and assumptions. In this section, however, we will focus only on those contributions that are closely related to our work. Stochastic models for the generation of workers' answers have been previously considered in \cite{EM-1}, \cite{EM-3}, \cite{EM-2}, \cite{caltech}. In \cite{EM-1} the parameters of the model are the per-worker confusion matrices that are jointly estimated with the ground truth answers. This method was extended in \cite{zhou2012learning}, where a distinct probability distribution was given over the answers for every task-worker pair. But \cite{zhou2012learning} does not consider a per-task parameter representing task difficulty. In our model, the distribution over answers for a task explicitly depends on the task difficulty that is essentially ``perceived" by workers attempting to solve it. More specifically, we explicitly model the \emph{interplay} between worker skill and task difficulty. Our focus in this paper has been mostly on crowdsourcing tasks that require an expertise (skill) present only in the minority of the crowd. In \cite{EM-2} and \cite{caltech}, task difficulty was considered explicitly, but only for the binary case. \cite{lakshminarayanan2013inferring} considered task difficulty for ordinal data, wherein a higher difficulty level adds more variance to the distribution of workers' elicited answers without affecting the mean (which equals the ground truth value). Our method of incorporating task difficulties is novel, as we use them in a comparative paradigm with worker skill in our soft threshold-based model.  Unlike \cite{caltech}, which assumes all tasks are drawn from the same classification domain, our approach is applicable even when the batch of tasks is \emph{heterogeneous}, \ie not necessarily all drawn from the same (classification) domain. This is due to the fact that our model does not consider domain-dependent features \cite{caltech} and also because our model is invariant to task-dependent permutations applied to the answer space. Note that \cite{raykar2012eliminating} allows biasing towards certain answers, which only applies when all the tasks belong to the same classification domain.

Adversarial workers in the binary case were accounted for in \cite{EM-2} and \cite{Karger11_NIPS}. In this work, we characterized adversarial behavior for a more generalized (multicategory) setting and proposed several realistic adversarial models. We also showed how we can retain the interpretation of negative weights as representing adversaries in generality from the binary \cite{Karger11_NIPS} to the multicategory case. Moreover, we showed that our approach exploits responses from (simple) adversaries to actually improve the overall performance. \cite{Karger11_NIPS} and \cite{karger2011budget} consider other statistical methods, such as correlation-based rules and low rank approximation of matrices. These methods have been studied for binary classification tasks. Our objective-based approach generalizes the weighted majority theme of these papers to a multicategory case, incorporating honest workers, adversaries, and spammers. We note that recently, \cite{cmu} have tried to extend the low rank approximation approach to the case when the tasks do not have a ground truth answer and the answers (from a categorical space) can be subjective. In this case, ``schools of thought" are discovered via clustering and the average size of clusters for each task is representative of its ease (clarity).

\section{Conclusion and Future Work}
This paper develops two novel paradigms to approach crowd aggregation and inference in multicategory crowdsourcing tasks: a generative stochastic modeling-based approach that explicitly models worker skill, intention and task difficulty and an objective-based non-generative approach that extends the notion of plurality voting to weighted plurality and also incorporates worker's intention. These approaches are specifically designed to overcome ``tyranny of the masses" by identifying the highly skilled workers in the crowd. We dissect a worker's accuracy into skill and intention, which allows us to extend the notion of adversarial behavior to a multiclass scenario. Along with experiments on synthetic and real world data, we evaluated our schemes using a heterogeneous crowd of simulated workers, where each worker was an automated classifier generated via boosting. From our experiments, we verified that our proposed schemes outperform other benchmark schemes, especially in the case when a small minority of highly skilled workers is present in a crowd of mostly low skilled workers. Moreover, our schemes were able to detect and leverage adversarial behavior. In the future, we would like to comprehensively evaluate more complex adversarial models, including the one proposed in this paper. We would also like to explore the possibility of different attacks on crowdsourcing systems, for instance collusion attacks, where a group of adversarial workers collude and submit the same (but incorrect) answer for a task. This may severely affect the performance of unsupervised methods, depending on the size of the group. We will also study the relationship between our objective-based schemes and low rank approximation schemes and attempt to extend message passing schemes to a multicategory setting.

\bibliographystyle{plain}

\begin{thebibliography}{10}

\bibitem{UCI}
K.~Bache and M.~Lichman.
\newblock {UCI} machine learning repository.
\newblock {\em University of California, Irvine, School of Information and
  Computer Sciences}, 2013.

\bibitem{benbouzid2012multiboost}
D.~Benbouzid, R.~Busa-Fekete, N.~Casagrande, F.D. Collin, B.~K{\'e}gl, et~al.
\newblock Multiboost: a multi-purpose boosting package.
\newblock {\em Journal of Machine Learning Research}, 13:549--553, 2012.

\bibitem{EM-1}
A.P. Dawid and A.M. Skene.
\newblock Maximum likelihood estimation of observer error-rates using the {EM}
  algorithm.
\newblock {\em Applied Statistics}, pages 20--28, 1979.

\bibitem{Original-EM}
A.P. Dempster, N.M. Laird, and D.B. Rubin.
\newblock Maximum likelihood from incomplete data via the {EM} algorithm.
\newblock {\em Journal of the Royal Statistical Society Series B
  (Methodological)}, pages 1--38, 1977.

\bibitem{douceur2002sybil}
J.~R. Douceur.
\newblock The sybil attack.
\newblock In {\em Peer-to-peer Systems}, pages 251--260. 2002.

\bibitem{feldman2004free}
M.~Feldman, C.~Papadimitriou, J.~Chuang, and I.~Stoica.
\newblock Free-riding and whitewashing in peer-to-peer systems.
\newblock In {\em ACM SIGCOMM Workshop on Practice and Theory of Incentives in
  Networked Systems}, pages 228--236, 2004.

\bibitem{graham2006unsupervised}
M.~W. Graham and D.~J. Miller.
\newblock Unsupervised learning of parsimonious mixtures on large spaces with
  integrated feature and component selection.
\newblock {\em IEEE Transactions on Signal Processing}, 54(4):1289--1303, 2006.

\bibitem{karger2011budget}
D.~R. Karger, S.~Oh, and D.~Shah.
\newblock Budget-optimal crowdsourcing using low-rank matrix approximations.
\newblock In {\em 49th IEEE Annual Allerton Conference on Communication,
  Control, and Computing}, pages 284--291, 2011.

\bibitem{Karger11_NIPS}
D.R. Karger, S.~Oh, and D.~Shah.
\newblock Iterative learning for reliable crowdsourcing systems.
\newblock {\em Advances in Neural Information Processing Systems}, 2011.

\bibitem{lakshminarayanan2013inferring}
B.~Lakshminarayanan and Y.~W. Teh.
\newblock Inferring ground truth from multi-annotator ordinal data: a
  probabilistic approach.
\newblock {\em arXiv.org preprint arXiv:1305.0015}, 2013.

\bibitem{meng1997algorithm}
X.~L. Meng and D.~Van~Dyk.
\newblock The {EM} algorithm—an old folk-song sung to a fast new tune.
\newblock {\em Journal of the Royal Statistical Society: Series B (Statistical
  Methodology)}, 59(3):511--567, 1997.

\bibitem{raykar2012eliminating}
V.~C. Raykar and S.~Yu.
\newblock Eliminating spammers and ranking annotators for crowdsourced labeling
  tasks.
\newblock {\em Journal of Machine Learning Research}, 13:491--518, 2012.

\bibitem{EM-3}
V.C. Raykar, S.~Yu, L.H. Zhao, G.H. Valadez, C.~Florin, L.~Bogoni, and L.~Moy.
\newblock Learning from crowds.
\newblock {\em Journal of Machine Learning Research}, 11:1297--1322, 2010.

\bibitem{schapire1990strength}
R.~E. Schapire.
\newblock The strength of weak learnability.
\newblock {\em Machine learning}, 5(2):197--227, 1990.

\bibitem{taskar1}
U.~Syed and B.~Taskar.
\newblock Semi-supervised learning with adversarially missing label
  information.
\newblock {\em Advances in Neural Information Processing Systems}, 2010.

\bibitem{szarvas2006multilingual}
G.~Szarvas, R.~Farkas, and A.~Kocsor.
\newblock A multilingual named entity recognition system using boosting and c4.
  5 decision tree learning algorithms.
\newblock {\em Discovery Science}, pages 267--278, 2006.

\bibitem{cmu}
Y.~Tian and J.~Zhu.
\newblock Learning from crowds in the presence of schools of thought.
\newblock In {\em ACM SIGKDD International Conference on Knowledge Discovery
  and Data Mining}, pages 226--234, 2012.

\bibitem{vukovic2009crowdsourcing}
M.~Vukovic.
\newblock Crowdsourcing for enterprises.
\newblock In {\em IEEE World Conference on Services-I}, pages 686--692, 2009.

\bibitem{caltech}
P.~Welinder, S.~Branson, S.~Belongie, and P.~Perona.
\newblock The multidimensional wisdom of crowds.
\newblock {\em Advances in Neural Information Processing Systems}, 6(7):8,
  2010.

\bibitem{pasadena}
P.~Welinder and P.~Perona.
\newblock Online crowdsourcing: rating annotators and obtaining cost-effective
  labels.
\newblock In {\em Proc. IEEE Conf on Computer Vision and Pattern Recognition
  Workshops (CVPRW)}, pages 25--32, 2010.

\bibitem{EM-2}
J.~Whitehill, P.~Ruvolo, T.~Wu, J.~Bergsma, and J.~Movellan.
\newblock Whose vote should count more: Optimal integration of labels from
  labelers of unknown expertise.
\newblock {\em Advances in Neural Information Processing Systems},
  22:2035--2043, 2009.

\bibitem{zhou2012learning}
D.~Zhou, J.~Platt, S.~Basu, and Y.~Mao.
\newblock Learning from the wisdom of crowds by minimax entropy.
\newblock {\em Advances in Neural Information Processing Systems}, pages
  2204--2212, 2012.

\end{thebibliography}

\end{document}